\documentclass[12pt,a4paper] {article}
\usepackage[dvips]{epsfig}
\usepackage{amsmath}

\begin{document}
\setlength{\unitlength}{1mm}

\def  \up      {\uparrow}
\def  \down    {\downarrow}
\def  \righ    {\rightarrow}
\def  \half    {\hbox{$\frac12$}}
\def  \ts      {\thinspace}
\def  \mt      {\mbox{$m_t$}}
\def  \ttbar   {\mbox{$t \bar t$}}
\def  \Wg      {\sl W-g}

\begin{titlepage}
\bigskip\bigskip\bigskip\bigskip\bigskip

\begin{center}{\large\bf
Spin Effects in Processes of Single Top Quark Production\\ at
Hadron Colliders}
\end{center}
\bigskip\bigskip

\begin{center}
E.~E.~Boos\footnote[1]{e-mail: boos@theory.sinp.msu.ru} and
A.~V.~Sherstnev\footnote[2]{e-mail: sherstnv@theory.sinp.msu.ru}
\bigskip

Skobeltsyn Institute of Nuclear Physics, Moscow State
University,\\ 119992 Moscow, Russia
\end{center}
\bigskip\bigskip\bigskip

\begin{center}{\large  Abstract}\end{center}

\noindent
We investigate spin properties of single top quark
production at hadron colliders. Based on an analogy with  
single top production and polarized top quark decay, we reproduce 
in a simple way the results by G.~Mahlon
and S.~Parke on the existence of preferred axes for the decomposition
of the top quark spin. For the $W^*$- and {\Wg}-fusion production modes 
these axes are related to the down-type quark momentum. 
The proposed method allows finding
kinematical conditions for the observation of top quark polarization
in a third process that contributes to single top production
and is important at LHC energies, 
the $tW$-process, in which spin effects are
smeared out by the contribution of diagrams with a QCD
$gt\bar{t}$-vertex. A simple Monte-Carlo analysis of spin
correlations for the $tW$-process with subsequent top decay is
given as an illustration.
\end{titlepage}

\section{Introduction}
The top quark is the heaviest Standard Model (SM) particle found so
far, with a mass $m_t{\sim}175$ GeV$\sim{v_H}/\sqrt{2}$ ($v_H$ -
vacuum expectation of Higgs field) and with a Yukawa coupling
very close to unity. This fact is probably related to a nature of
the electroweak symmetry-breaking mechanism. In the SM the top quark
is very heavy but at the same time is assumed to be point-like.
Because of
these and other unusual top quark properties, possible
deviations from SM predictions might be first manifest in
the top quark sector.

As a consequence of a large quark mass the top quark's electroweak decay
$t{\righ}W^-b$ (in the framework of SM) proceeds so rapidly that
hadron bound states do not have enough time to form \cite{Bigi}. This
leads to the fact that angular distributions of top quark decay
products are mainly determined by the momentum and spin state of
the $t$-quark itself and are not smeared out by hadronization
effects~\cite{Kuhn-1}.

Top quarks being produced singly through the electroweak
interaction give a unique opportunity to investigate a number of
delicate top quark properties. In particular, single top production is the only
source of a direct measurement of the CKM matrix element $V_{tb}$.
Because of the large production rate at hadron colliders,
single $t$-quark production provides an important background to
various processes expected in the SM and beyond, such as
Higgs boson or SUSY particle production. Single $t$-quarks are
expected to be produced with a high degree of polarization because
of the pure (V--A)-structure of the production $Wtb$-vertex assumed
in the SM. Spin properties of the all single top
production processes are thus of special interest.

G.~Mahlon and S.~Parke have found that the direction of the spin
of the single top quark in the production processes of $W^*$ and {\Wg} fusion 
coincides
with the momentum of the down-type quark. For the $W^*$-process the
down-type quark is the $\bar{d}$-quark in the initial  
state~\cite{MahlonParkeTeva},
while in  {\Wg} fusion it is either the final spectator quark
(in the majority of events), or the $\overline{d}$-quark in the initial
state~\cite{MahlonParkeTeva}, \cite{MahlonParkeLHC}.

In this paper we show that the above results have a simple
explanation if one considers single top production processes as
decays of a polarized top quark considered backwards in time. 
By considering the analogy with polarized top decay,
we find conditions for significant
$t$-quark polarization in the third-most important single top production 
process at LHC energies, the so-called $tW$-process.

This paper is organized as follows. In Section 2 we mention
briefly all three processes for single $t$-quark production at
hadron colliders and  recall the results obtained by G.~Mahlon and
S.~Parke for the $W^*$-process and {\Wg} fusion for top quark polarization.
In Section 3 we 
give a simple explanation of these results based on properties of 
polarized $t$-quark decay and  
analyze spin polarization for the more
complicated $tW$-process of single top production  including
subsequent top decay. A Monte-Carlo study of the $t$-quark spin
effects in the $tW$-process is given in Section 4. The conclusions are
presented in Section 5. 

\section{Single Top Quark Production Processes}

\begin{figure}
\begin{center}
  \epsffile{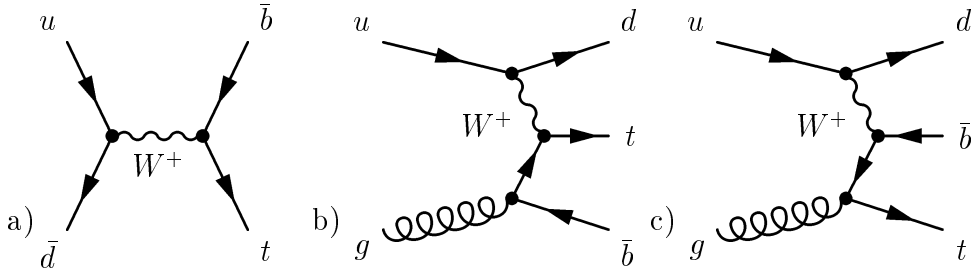}
 \caption{Typical diagrams for the $W^*$-process (a)
          and {\Wg} fusion (b-c).}
 \label{WG}
\end{center}
\end{figure}

\begin{figure}
\begin{center}
 \epsffile{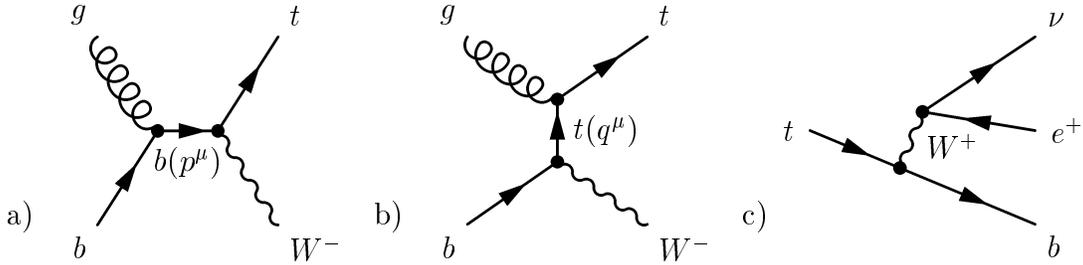}
 \caption{Diagrams for the $tW$-process (a-b) and LO $t$-quark decay (c).}
 \label{tW}
\end{center}
\end{figure}

There are three SM processes of single top quark production at
hadron colliders\footnote{See the complete set of contributing
parton subprocesses in~\cite{BoosBelyaev-1}.}; some representative
diagrams are shown in Figures~\ref{WG} and~\ref{tW}(a-b)\footnote{All
Feynman diagrams in the paper were made with the help of the latex
package feynmf written by T.~Ohl \cite{FeynmanDiag}.}.
Each of the processes may be characterized by the virtuality
$Q^2_W$, the four-momentum squared of the participating $W$-boson:
\begin{itemize}
\item $t$-channel $W$-exchange ($Q^2_W<0$): the characteristic diagrams
are
depicted in Fig.~\ref{WG}(b-c). This process has the largest cross
section both at the Tevatron and LHC. It is referred as {\Wg}
fusion for the $2\righ{}3$ diagrams shown, as well as for the
$2\righ{}2$ part with the $b$-quark in the initial state because the
initial $b$-quark ultimately arises from a gluon splitting $g\righ
b\bar{b}$. We call the diagram~\ref{WG}(b) the $Wtb$-diagram
because the top quark is produced at the $Wtb$-vertex. The diagram
\ref{WG}(c) is called the $gtt$-diagram. One should
stress that for the $2\righ{}2$ process there is only one
$Wtb$-diagram.

\item $s$-channel $W$-exchange ($Q^2_W>0$): the $W^*$-process.
The characteristic diagram is depicted in Fig.~\ref{WG}(a). This
process has a predicted rate for Run 2 at the Tevatron only about 2.5 times
smaller than the {\Wg} fusion rate. Although the process should be
observable at the LHC, it has a cross section about 25 times
smaller than the {\Wg} fusion one (see the Table).

\item Real $W$ production ($Q^2_W=m_W^2$): $tW$-process. A single
top quark appears in association with a real $W$-boson as shown in
Fig.~\ref{tW}(a-b). These diagrams are also called the $Wtb$-diagram
and $gtt$-diagram, depending on the type of the $t$-quark production
vertex. This process has a very small production cross section at
the Tevatron because of two massive particles in the final state, while
at the LHC the rate is significant. We point out the presence of
two $Wtb$- and $gtt$-diagrams even for the $2\righ{}2$ part of the
$tW$-process. As it will be shown this fact leads to additional
complications for the analysis of the top quark spin properties.
\end{itemize}

The basic cross sections have been calculated to the NLO level in
Ref.~\cite{Willenbrock1} for {\Wg} fusion and in
Ref.~\cite{Willenbrock2} for the $W^*$-process, and to LO for the
$tW$-process in Ref.~\cite{Tait, BoosBelyaev-2} (see the Table).

\begin{table}
\begin{center}
\caption[fake]{Total cross sections of single $t$-quark production 
processes for \mt=175 $\pm 2$~GeV~\cite{MainLHCTop}.} \vspace*{0.3cm}
\begin{tabular}{|l|ccc|}
\hline
  process      & {\Wg} fusion  & $W^*$-process  & $tW$-process \\
\hline
  LHC (pb)     & $245\pm 27$ &  $10.2\pm 0.7$ & $62.0+16.6/-3.6$ \\
\hline
  Tevatron (pb)& $2.12\pm 0.1$& $0.88\pm 0.05$& $0.093\pm 0.024$ \\
\hline
\end{tabular}\label{tab:sigma}
\end{center}
\end{table}
Because of the unique (V--A) structure of the $Wtb$-vertex in the SM
the electroweak single top production processes have very
interesting top spin properties~\cite{MahlonParkeTeva, MahlonParkeLHC,YuanTait}.
For the case of $W^*$- and {\Wg} fusion, 
G.~Mahlon and S.~Parke~\cite{MahlonParkeTeva, MahlonParkeLHC} 
have found rather compact 
formulae for single top production 
using a  convenient formalism for top quark spin
with an arbitrary spin direction. For the $W^*$-process
it was shown the expressions for the squared matrix elements
$|M(+)|^2$ $(|M(-)|^2)$ for the top quark polarized along (opposite) some 
direction $\vec{n}_t$ in its rest frame have a simple form 
\footnote{As in
~\cite{MahlonParkeTeva, MahlonParkeLHC} we denote the momentum
of the each particle by its symbol.}

$$
|M(+,u\bar{d}\righ{}t\bar{b})|^2= 9g_W^4 |V_{ud}V_{tb}|^2
\frac{(2d\cdot t_-)(2u\cdot b)}{(w^2-m_W^2)^2+(m_W\Gamma_W)^2}
\label{WStarMatrix+}
$$
for the spin up top quark and
$$
|M(-,u\bar{d}\righ{}t\bar{b})|^2= 9g_W^4 |V_{ud}V_{tb}|^2
\frac{(2d\cdot t_+)(2u\cdot b)}{(w^2-m_W^2)^2+(m_W\Gamma_W)^2}
\label{WStarMatrix-}
$$
for the spin down top quark,
where $t_{+}=\half(t+m_ts)$ and $t_{-}=\half(t-m_ts)$. 
Spin vector of the top quark $s^{\mu}$ has a form  
$s^{\mu}=(0,\vec{n}_t)$ in the $t$-quark rest frame.

From these two formulas one can see that if one takes the
polarization vector $\vec{n}_t$  along the
direction of the $\bar{d}$-quark three-vector momentum
$\vec{p}_{\bar{d}}{}^*$ in the top rest frame,
$$\vec{n}_{\bar{d}}=\vec{p}_{\bar{d}}{}^*/p_{\bar{d}}{}^*,$$
the squared matrix element $|M(-,u\bar{d}\righ{}t\bar{b})|^2$
exactly equals zero.
G.~Mahlon and S.~Parke have interpreted this result in terms of the
direction of the $t$-quark spin coinciding with the direction of the
$\bar{d}$-quark momentum ($\vec{s}_t\up\up\vec{p}_{\bar{d}}{}^*$). At the
Tevatron the largest contribution to the total cross section comes
from the case where the $\bar{d}$-quarks come from the
antiproton. The best choice of the spin decomposition axis is thus the
antiproton beam, which the authors call the ``antiproton'' spin
basis.

The situation in the case of {\Wg} fusion process is more complex.
It was mentioned already long ago in the paper by Willenbrock and
Dicus~\cite{WillenbrockDicus} that the {\Wg} fusion process is
dominated by the configuration where the $\bar{b}$ quark from
$g\righ b\bar{b}$ splitting is nearly collinear with the incoming
gluon, leading to a logarithmic factor $\ln(m_t^2/m_b^2)$ in the
total cross section. It is well known that these large corrections
are resummed, being absorbed into the $b$-quark parton distribution
function.  The correct LO rate is obtained then by summing up
contributions of the $2\righ 2$ process $ub\righ td$  and the
 $2\righ 3$ process mentioned above, with the subtraction of the first
$g\righ b\bar{b}$ splitting term in order to avoid double
counting. The spin properties of the top quark in the $2\righ 2$ process,
both in the subtracted term
and in the dominant contribution of the $2\righ 3$
process coming from the $Wtb$-diagram, are similar in the top rest
frame \cite{MahlonParkeLHC}. 
In all these cases the top is
produced at the $Wtb$-vertex. The most effective spin decomposition
axis is once more the direction $\vec{n}_d{}^*$.
But the problem of extracting the d-quark momentum direction is less
obvious now, because the d-quark appears both in the initial and
final states. Two spin bases have been introduced here, the
``spectator jet'' basis related to the dominant single top
contribution with the d-quark in the final state for the Tevatron
and LHC, and the ``$\eta$-beamline'' basis related to the d-quark in
the initial state. The last is important for an analysis of $\bar{t}$-quark
spin properties at the LHC.

In all cases the main spin effects are manifested in the simplest
$2\righ 2$ processes with the b-quark in the initial state.

One remark is in order here. The fact that the top quark is polarized in 
the direction of $\vec{n}_d{}^*$ means that
$|M(+)|^2\ne 0$ and $|M(-)|^2 = 0$.
For the $2\righ 3$ diagrams of the {\Wg}-fusion shown in Fig.~\ref{WG}(b-c) 
the value of $|M(+)|^2$ comes from the contributions of all three parts,
the $Wtb$, the $gtt$-diagrams and the interference between them.
The $Wtb$-diagram and the interference do not 
contribute to $|M(-)|^2$. Indeed, one can show that if 
$\vec{n}=\vec{n}_d{}^*$ then 
$|M_{Wtb}(-,ug\righ{}t\bar{b}d)|^2=0$
and the interference $M_{Wtb}(-)*M_{gtt}^*(-)+h.c.$ is also exactly equal 
to zero. However,  the $gtt$-diagram in general does not lead to any 
preferable direction of the top polarization, and gives nonzero  contributions 
to both $|M(+)|^2$ and $|M(-)|^2$.  
Fortunately, $|M(-)|^2=|M_{gtt}(-,ug\righ{}t\bar{b}d)|^2$ is much
smaller than the complete $|M(+)|^2$, which comes from the large $Wtb$ squared
diagram and  the interference. 
One can thus neglect the $|M(-)|^2$ part and conclude that
with a good accuracy the top quark is polarized along the d-quark momentum in 
its rest frame.

For the $tW$ mechanism of single top production the situation is
similar in a sense that once more large corrections from the soft
b-quark region are resummed by introducing the b-quark distribution
function. This $2\righ 2$ contribution with the b-quark in the initial
state dominates the production rate; top spin
properties could therefore be studied by considering the $2\righ 2$ process.
However, in contrast to the {\Wg} fusion case, the $gtt$-diagram
contribution to $tW$-process is significant and the corresponding
diagram appears already in the $2\righ 2$ process as seen from the
Fig.~\ref{tW}(a-b).
The situation with top polarization in the $tW$-process
thus requires special consideration.

\section{Top Quark Polarization in the $tW$-process}
\label{WgW*andDecay}

In order to understand top quark spin properties in the $tW$-process it
is useful to look at the results mentioned above for the $W^*$ and
{\Wg} fusion processes from a different point of view.

The results on strong correlations between the $t$-quark spin and
d-quark momentum can be obtained and explained in a very simple
way based on the properties of polarized top decay. The diagram in
Fig.~\ref{tW}(c) for the LO $t$-quark decay is topologically
equivalent to the diagram for $W^*$-process and to the $2\righ 2$ part
($ub\righ td$) of the {\Wg} fusion (the latter reproduces the main
features of the top spin properties in the complete $W$-gluon
fusion as we mentioned). The decay matrix element is the same as that
for production,  and these processes differ only in the
kinematical region over which they are integrated to get the decay
width and the cross section. So the two single top production
processes can be considered as a corresponding top decay backwards in
time.  The analogue of the d-quark in top decay $t\righ{}l^+\nu_lb$ is the
charged lepton, since in the SM both the d-quark and the charged
lepton are the down components of the electroweak doublets.
Therefore the properties of the charged lepton from
the decay should be similar to those of the d-quark.

The decays of polarized $t$-quarks have been investigated
comprehensively many times both in LO and NLO~\cite{Kuhn-1},
\cite{Kuhn-2, Kuhn-3,TDecayLoop-1, TDecayLoop-2}. The main result concerning the
positron is that it is the best spin probe of the $t$-quark
polarization~\cite{Kuhn-2}. The angular distribution of the
positron from the decay of the top quark polarized along some axis
$\vec{n}$ in the $t$-quark rest frame is
$$\frac1\Gamma\;\frac{d\Gamma(\vec{s}_{t}\up\up \vec{n})}
{d\cos{\theta}}=\frac{1+\cos{\theta}}2\;,$$ where $\theta$ is the
angle between the $t$-quark spin direction $\vec{n}$ and the momentum of
the positron. So one gets unity for the distribution if the
direction of the top polarization  coincides with the positron
momentum direction ($\theta=0$) and zero for the opposite
direction. In other words this means that for an arbitrary top
polarization state along some direction, only the state with the
projection along the positron momentum direction contributes to the
decay.

One should stress that the existence of a preferred spin axis is
related with the (V--A)-structure of the $Wtb$-vertex. In 
view of this fact a separation of one $t$-quark spin state does not
cause astonishment. In some sense this situation is similar
to the situation with the neutrino. A
chiral projector $P_L$ cuts out one helicity state of the
neutrino. The top quark is the massive particle and its helicity and
chirality states surely do not coincide. However, for the top the
$Wtb$-vertex with exact (V--A)-structure leaves only the one spin
projector of the $t$-quark in the non-polarized spin density matrix
$P=p^{\mu}\gamma_{\mu}+m$. The interpretation
is easiest in the $t$-quark rest frame, where $\vec{s}_{t}\up\up
\vec{n}_{e^+}{}^*$.

Now returning back to the $2\righ 2$ production processes like
$u\bar{d}\righ t\bar{b}$ ($W^*$) or $ub\righ td$ ($W-g$ fusion), and 
keeping in mind the
analogy between the positron and the d-quark one can immediately
conclude that the top quark is produced in a definite spin state
with the top spin direction in its rest frame being along the
$\bar{d}$ or $d$ quark momentum direction.

Also it becomes obvious that after top production with a
definite polarization state only the projection of its spin state
along the axis corresponding to the positron momentum as seen from 
the top
rest frame contribute. Therefore the overall matrix element
squared, including top production and decay, is proportional to
$1+\cos{\theta_{e^+,d}^*}$, where $\theta_{e^+,d}^*$ is the angle
between the momenta of the positron and the $d$ (or $\bar{d}$)-quark in the
top rest frame. The corresponding distribution has the form:
$$
\frac1\sigma\frac{d\sigma}{d\cos{\theta{}^*_{e^+,d}}}=
\frac{1+\cos{\theta_{e^+,d}^*}}2
$$
This result exactly reproduces the result by G.~Mahlon and
S.~Parke discussed in the previous section, and corresponds to the
highest possible spin correlation between production and decay.
Correspondingly, the best variable for experimental analysis of
the $t$-quark spin properties is $\cos{\theta_{e^+,d}^*}$.

Although the $tW$ process, like the $W^*$-process or the 
{\Wg} fusion part of the
process with a b-quark in the initial state, is of the
$2\righ 2$ type, there are two diagrams (Fig.~\ref{tW}a-b) and one of
them is of the $gtt$-type. As we mentioned, in contrast to the
{\Wg} fusion the contributions of both these two diagrams are
comparable in rate. One can not simply remove one of these two
diagrams since only together do they form a gauge-invariant
set of diagrams.

\begin{figure}
\begin{center}
 \epsffile{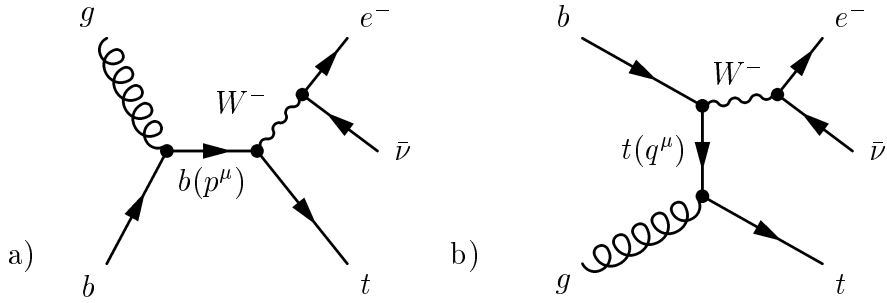}
\caption{ Diagrams of $tW$-process with decay of W-boson
$W^-\righ{e^+\nu}$} \label{Real-tW-process}
\end{center}
\end{figure}

Let us consider how the analogy with the top decay helps in this
case. If we add to the production diagram the leptonic decay of
the $W$-boson produced in association with a top quark we get
the diagrams presented in Fig.~\ref{Real-tW-process}.
\begin{figure}
\begin{center}
 \epsffile{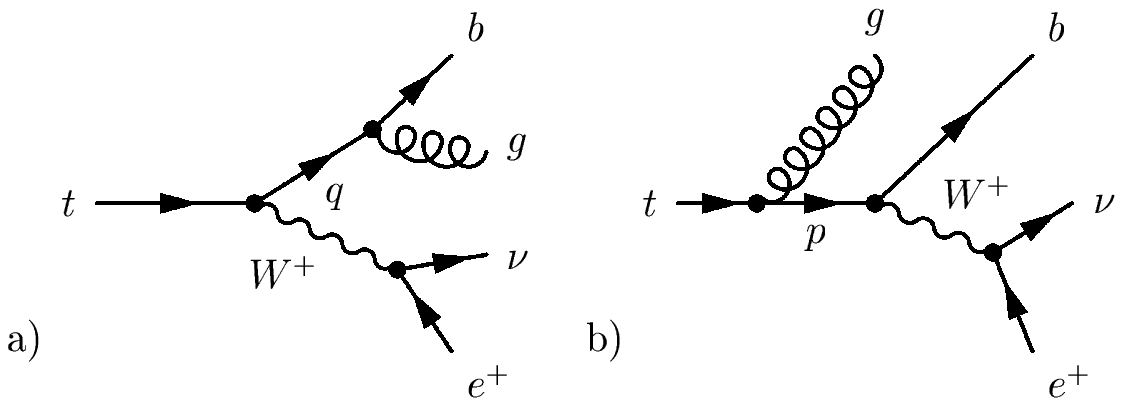}
 \caption{ Diagrams of the decay $t\righ b\ts e^+\nu+g$}
\label{t-beNG}
\end{center}
\end{figure}
The diagrams of the $tW$-process with $W$-boson decay in Fig.~\ref{Real-tW-process} are
topologically equivalent to the diagrams of
top quark decay with radiation of an additional gluon $t\righ{}be^+\nu+g$ (Fig.\ref{t-beNG}), 
which is
simpler to analyze. Let the axis for the top spin decomposition be the
direction of the positron momentum produced in the $W$-boson
decay, $\vec{n}_e{}^*$. Then one can easily check that the
contribution of the $Wtb$-diagram squared to the top quark spin down
configuration $|M_{Wtb}(-,t\righ be^+\nu+g)|^2$ and the contribution of the
$Wtb$--$gtt$  interference diagrams $|M_{Wtb*gtt}(-,t\righ
be^+\nu+g)|$ equal to zero exactly in the same manner as it was for
the $W^*$-process. However, the contribution of the $gtt$-diagram squared 
$|M_{gtt}(-,t\righ be^+\nu+g)|^2$ does not vanish.  
It turns out that the contribution for polarized
$t$-quark decay has a very simple symbolic structure
$$
|M_{gtt}(-,t\righ be^+\nu+g)|^2= \frac{\displaystyle 2\ts g^2_s\ts
g^2_W |V_{tb}|^2}{\displaystyle (w^2-m_W^2)^2+(m_W\Gamma_W)^2}\ts
\frac{\displaystyle 1}{\displaystyle (t\cdot g)^2}\ts(b\cdot\nu)
\Bigl[m_tE^*_eE^{*2}_g(1-\bar{n_e^*}\bar{n_g^*})^2\Bigr]\;,
$$
where $E^*_e$ and $E^*_g$ are the energies of the positron and the
gluon, and $\vec{n}_e{}^*$ and $\vec{n}_g{}^*$ are the directions
of their 3-momenta in the top rest frame, respectively. However if
the direction of the emitted gluon gets closer to the positron
direction in the top quark rest frame, or if the energy of the gluon
gets closer to zero, the $gtt$-diagram contribution, where
$\vec{s}_t\up\down\vec{p}_e{}^*$, goes to zero as well. In the
limit $\cos{\theta_{e^-,g}^*}\rightarrow 0$ the top quark will be
produced with a definite spin direction along the positron
momentum.

Now let us turn to the $tW$ production process and the analogy
with the radiative decay. From the above considerations it follows
that in single top production in association with a $W$-boson (see
Fig.~\ref{Real-tW-process}) the top quark spin in its rest frame
approaches $100\%$ polarization along the momentum of the produced
electron in the case when the electron momentum approaches the
direction of the gluon. The same happens in the limit of zero gluon
energy.

To get the cleanest top spin state, one should thus require the variable
$\cos{\theta_{e^-,g}^*}$ to be as close as possible to unity.
One can also apply an 
upper bound on the gluon energy in the top rest frame. We have found
from MC studies that an excellent  variable to cut on is a combination of
these two quantities, $X_{g,e^-}=E^*_g(1-\cos{\theta^*_{e^-,g}})$.

Now if one considers the subsequent top quark decay we know
already that only those projections of the top spin state
contribute to the top decay which coincide to the direction of
the positron from the top decay. Therefore from our considerations
it follows that a good variable to probe spin properties of the
top quark in the $tW$-process is the $\cos{\theta_{e^-,e^+}^*}$
where the electron comes from the $W$-boson decay
produced in association with top, and the positron comes from
the top decay.

\section{Monte-Carlo Results} \label{MonteCarlo}

There are 4 modes to search for the $tW$-process, depending on the
decay channels of the $t$-quark and $W^-$--boson. Taking into
account $t$-quark spin, all of them are not equivalent.

\begin{description}
\item[dilepton channel:] $t$-quark ($W^+$ boson from top decay) and
$W^-$-boson decay to leptons. One can determine easily the directions
of the $e^+$ and $e^-$ momenta. However, finding the $t$-quark rest frame is a
problem because of the two neutrinos in the final state. If the 6
components of the neutrino momenta are introduced as unknown
variables one gets only 5 equations: 2 equations from the measured
missing $P_T$ vector, and 3 equations from the 3 known masses,
$M_t$ and two $M_W$. An additional condition can be obtained from the
simultaneous measurement of the semileptonic mode with 
hadronic top decay, and consequently a normalization of the leptonic
mode. However the accuracy of such a procedure needs to be studied
in detail \cite{BoosSherstnev}. Also the production rate includes
two small branching ratios $Br(W\righ{l\nu_l})=2/9$ if $l$ is an
electron or muon, and $1/3$ if one includes $\tau$ in the analysis.
\item[t-leptonic channel:] top quark decays leptonically
and $W$-boson -- hadronically. Because of the analogy between the d-quark
and the electron discussed in the previous section, the variable used for
top spin analysis will be the angle between the positron from the top
decay and the d-quark from the $W$ decay. This mode also has a problem
of the reconstruction of the $t$-quark rest frame, since one should
extract the longitudinal part of the neutrino momentum solving the
equation for the $W$-boson mass.  Moreover, one needs to separate
the $d$-quark jet from the $u$-quark jet from the decay $W^-\righ
d\bar{u}$ which is very problematic.

\item[W-leptonic channel:] top quark decays hadronically and
$W$-boson -- leptonically. In contrast to the t-leptonic channel, in the
W-leptonic mode the $t$-quark rest
frame could be easily extracted, but the problem of a separation
of the $\bar{d}$-quark jet in the $t$-quark decay remains.
\item[hadronic channel:] Although this process does not have
troubles with the top and $W$ mass reconstruction the problem of $d$
and $\bar{d}$ jet separation from the corresponding $\bar{u}$ and
$u$ jet remains and leads to additional uncertainties. Also the
QCD background is very large in this case.
\end{description}

The complete MC analysis of the top spin properties in the $tW$-mode
is a complicated problem and it is beyond the scope of this paper. We
present here only some illustrative MC results for the dilepton
channel, ignoring for the moment the problems of the top rest frame,
reconstruction and contributions of backgrounds. The detail MC
analysis including backgrounds will be presented elsewhere
\cite{BoosSherstnev}.

We have used the program package CompHEP \cite{CompHEP} for the
Monte-Carlo generation of $tW$-process events. The events are
generated for a proton-proton initial state at the nominal LHC energy,
$\sqrt{s}=14$ TeV, using CTEQ5M1 structure functions \cite{CTEQ}.

\begin{figure}
\begin{center}
  \epsfxsize=12cm
  \epsfysize=8.5cm
  \epsffile{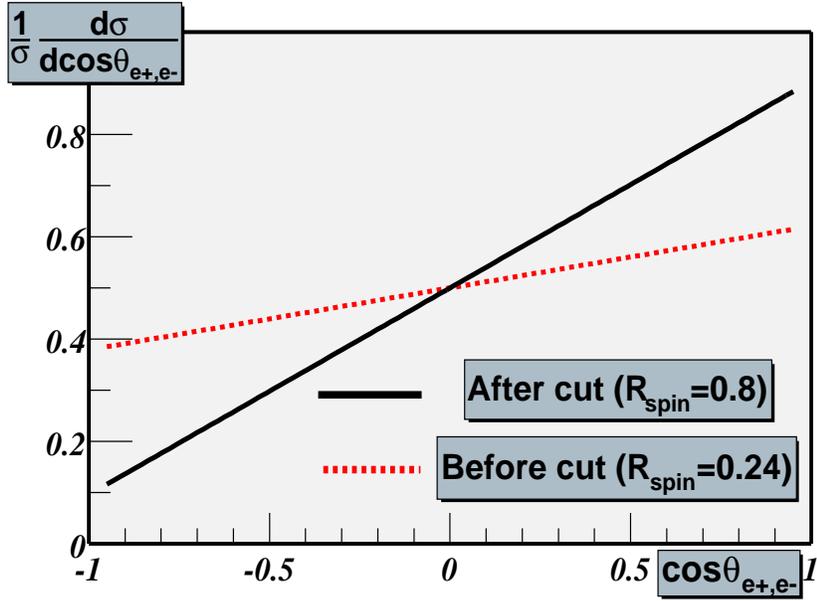}
\caption{The normalized distributions
$d\sigma/d\cos{\theta^*_{e^+,e^-}}$ for the $tW$-process: dashed line
-- without cuts, solid line -- cut $X_{g,e^-}<110$
GeV.}\label{cut}
\end{center}
\end{figure}

The normalized distribution $d\sigma/d\cos{\theta^*_{e^+,e^-}}$
for the $tW$-process without any cuts is shown in
Fig.~\ref{cut} (dashed line). As was explained above, if one applies
cuts on the variables
$\cos{\theta^*_{e^-,g}}=\vec{n}_g{}^*{\cdot}\vec{n}_{e^-}{}^*$ and
$E^*_g$ in the top quark rest frame the degree of $t$-quark polarization
of the event sample has to increase. Previously we
mentioned the best variable here is the combination of the angle
and gluon energy, of the form
$X_{g,e^-}=E^*_g(1-\cos{\theta_{e^-,g^*}})$. If the variable
$X_{g,e^-}$ is small enough, the $gtt$-diagram contribution is also
small. The solid curve in Fig.~\ref{cut} demonstrates a much higher
spin correlation after the cut $X_{g,e^-}<110$ GeV is applied.

\begin{figure}
\begin{center}
  \epsfxsize=12cm
  \epsfysize=8.5cm
  \epsffile{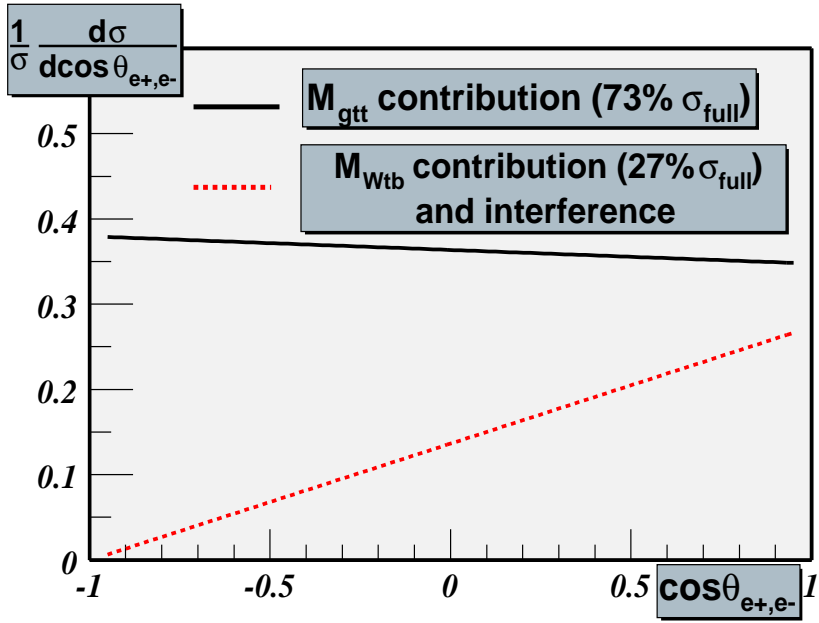}
\caption{The normalized distributions $d\sigma/d\cos{\theta^*_{e^+,e^-}}$ for
the $gtt$-diagram and the $Wtb$-diagram plus their interference.}
\label{divided}
\end{center}
\end{figure}
It is useful to analyze the spin asymmetry $R_{spin}$, which is
determined from the formula
$$\frac1\sigma\frac{d\sigma}{d\cos{\theta{}^*_{e^+,\vec{s}}}}=
\frac{1+R_{spin}(\vec{s}){\cdot}\cos{\theta^*_{e^+,\vec{s}}}}2$$
and changes between -1 to 1 for different spin axes. For a chosen
spin axis $\vec{s}$ along the electron momentum direction one can
express the value of the spin asymmetry  as
$$R_{spin}=\frac{y(1)-y(-1)}{y(1)+y(-1)}$$ where
$y(x)=d\sigma/d\cos{\theta^*_{e^+,e^-}}$ and
$x=\cos{\theta^*_{e^+,e^-}}$. 
If top quark would be fully polarized along the $n_{e^-}$ direction the spin asymmetry 
$R_{spin}$ would be equaled to 1. 

In the case of the $tW$-process the spin
asymmetry equals $R_{spin}\approx{0.24}$. The smallness of $R_{spin}$
is a result  of the influence of the $gtt$-diagram. The individual
diagram contributions in the t'Hooft-Feynman gauge are shown in
Fig.~\ref{divided}. Although the contribution of the $Wtb$-diagram
is large, the interference between $Wtb$- and $gtt$-diagrams is
destructive and leads to a cancellation in the angular
distribution. The $gtt$-diagram contribution is consequently the largest.

\begin{figure}
\begin{center}
  \epsfxsize=12cm
  \epsfysize=8.5cm
  \epsffile{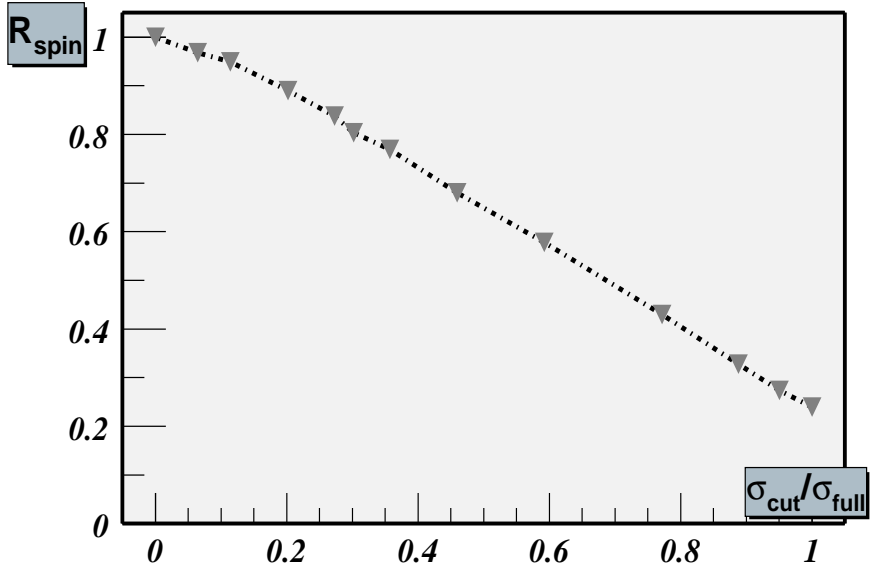}
\caption{The spin asymmetry $R_{spin}$ vs. the fall of the cross
section for the $tW$-process with the $X_{g,e^-}$ cut. } \label{dgtop}
\end{center}
\end{figure}
As it was explained above, in the limit $X_{g,e^-}\rightarrow 0$ the spin
asymmetry approaches 1. In Fig.~\ref{cut} the normalized
angular distribution demonstrates the increase of the spin asymmetry
from 0.24 to 0.8 if a proper cut on $X_{g,e^-}$ is applied. With
this cut $tW$-process cross section drops by a factor of 3. The
dependence of $R_{spin}$ on the cross section after the cut is shown in
Fig.~\ref{dgtop}.

\begin{figure}
\begin{center}
  \epsfxsize=12cm
  \epsfysize=8.5cm
  \epsffile{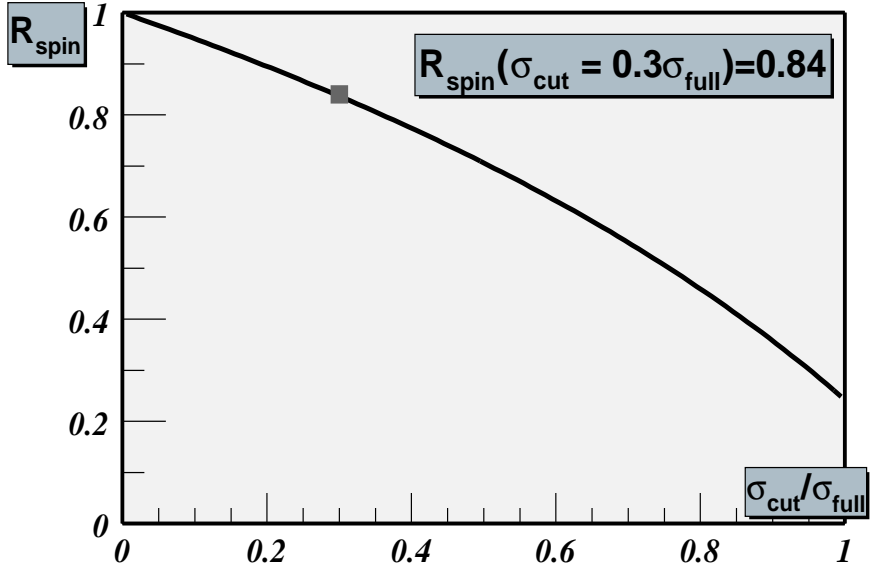}
\caption{The spin asymmetry $R_{spin}$ vs. the fall of the cross
section of $tW$-process with the $\cos{\theta_{e^-,e^+}^*}$ cut.}
\label{detop}
\end{center}
\end{figure}
In order to increase the spin asymmetry $R_{spin}$ one can also
use properties of the distribution
$d\sigma/d\cos{\theta^*_{e^+,e^-}}$ itself. 
From Fig.~\ref{divided} we can see that the distribution of events 
from the $Wtb$ squared diagram plus the interference has a peak near 
$\cos{\theta^*_{e^+,e^-}}=1$. These events correspond to 
$t$-quarks which are fully-polarized 
along the $e^-$ momentum. The distribution 
$d\sigma/d\cos{\theta^*_{e^+,e^-}}$ 
of events with the opposite top quark polarization
has a peak near $\cos{\theta^*_{e^+,e^-}}=-1$.
The contribution from the $gtt$ squared diagram is almost flat,
which means it gives about equal numbers of events with both top quark
polarizations. Therefore, if one cuts out the region of
$\cos{\theta^*_{e^+,e^-}}$ close to $-1$, a significant fraction of
the events with the opposite spin configuration will be removed. The
Fig.~\ref{detop} shows the increase of $R_{spin}$ depending on the
ratio of cross section after the $\cos{\theta^*_{e^+,e^-}}$ cut 
to the full one.

In practice in order to get an event sample with higher top
polarization one can use cuts on both variables  $X_{g,e^-}$ and
$\cos{\theta^*_{e^+,e^-}}$ simultaneously.

\section{Conclusions}

In the paper a close relation between spin properties in single
top quark production and in polarized top quark decay has been
pointed out. Based on the known fact that the positron in top
decay $t\righ be^+\nu_e$ is the best probe of the top spin, and
on the analogy between the positron and the down-type quark, one can
conclude the top quark spin in each event should follow the
direction of the down-type quark momentum in the top quark rest frame.
This is the direction of the initial $\bar{d}$-quark for the
$W^*$-process, and  the dominant direction of the final $d$-quark for the
{\Wg} fusion process. Also it becomes obvious that the best variable to
observe maximal top spin correlations between the top 
production and subsequent decay is the angle between this down-type
quark direction in  the production processes and the charged lepton
(or $d$-quark) direction from the top decay in the top rest frame.

Spin properties of the $t$-quark in a third production mechanism,
the $tW$-process, are more
complicated to analyze. In this case the diagram where the top
quark is produced at the (QCD) $gtt$-vertex contributes with a
significant rate in contrast to the two processes discussed earlier. The
observed analogy of the $tW$ production mode to radiative
polarized top decay allows finding  effective kinematical
variables, e.g. $X_{g,e^-}$ and $\cos{\theta^*_{e^+,e^-}}$, to reduce the
contribution with the opposite polarization. Cuts on
these variables select top quarks 
produced with the polarization vector preferentially close 
to the direction of the charged lepton or the $d$-quark momentum
from the associated $W$ decay. These cuts thus can raise the
observed spin
asymmetry $R_{spin}$ in the process. We present a Monte-Carlo
analysis only for the dilepton channel and at the parton level for
final states. More realistic Monte-Carlo generation taking into
account hadronization effects, identification of particles, and
other detailed effects is in progress.

\section{Acknowledgments}
The authors are grateful to G.~Belanger and F.~Boudjema for useful
discussions.  This research was supported in part by CERN-INTAS
grant 99-0377, INTAS grant 00-0679, and RFBR grant 01-02-16710.
E.B. thanks the Bessel Research Award of the Humboldt Foundation.

\end{document}